\newcounter{myctr}
\def\myitem{\refstepcounter{myctr}\bibfont\noindent\ifnum\themyctr>9\else\phantom{0}\fi\hangindent17pt\themyctr.\enskip}

\documentclass{ws-ijqi}
\usepackage{color}

\begin{document}

\markboth{Gian Luca Giorgi, Fernando Galve and Roberta Zambrini}
{Robustness of different indicators of quantumness in the presence of dissipation}

%%%%%%%%%%%%%%%%%%%%% Publisher's Area please ignore %%%%%%%%%%%%%%
\catchline{}{}{}{}{}
%%%%%%%%%%%%%%%%%%%%%%%%%%%%%%%%%%%%%%%%%%%%%%%%%%%%%%%%%%%%%%%%%%%

% \title{INSTRUCTIONS FOR TYPESETTING MANUSCRIPTS\\ 
% USING COMPUTER SOFTWARE\footnote{For the 
% title, try not to use more than 3 lines. 
% Typeset the title in 10~pt Times roman, uppercase and boldface.}  }
\title{ROBUSTNESS OF DIFFERENT INDICATORS OF QUANTUMNESS IN THE PRESENCE OF DISSIPATION}

\author{Gian Luca Giorgi}
\address{IFISC (UIB-CSIC),
Instituto de F\'isica Interdisciplinar y Sistemas Complejos, UIB Campus,
E-07122 Palma de Mallorca, Spain\\gianluca@ifisc.uib-csic.es}

\author{Fernando Galve}
\address{IFISC (UIB-CSIC),
Instituto de F\'isica Interdisciplinar y Sistemas Complejos, UIB Campus,
E-07122 Palma de Mallorca, Spain\\fernando@ifisc.uib-csic.es}

\author{Roberta Zambrini}\address{IFISC (UIB-CSIC),
Instituto de F\'isica Interdisciplinar y Sistemas Complejos, UIB Campus,
E-07122 Palma de Mallorca, Spain\\roberta@ifisc.uib-csic.es}
% \author{FIRST AUTHOR\footnote{Typeset names in
% 8~pt roman, uppercase. Use the footnote to indicate the
% present or permanent address of the author.}}
% 
% \address{University Department, University Name, Address\\
% City, State ZIP/Zone,
% Country\footnote{State completely without abbreviations, the 
% affiliation and mailing address, including country. Typeset in 
% 8~pt Times italic.}\\
% first\_author@university.edu}
% 
% \author{SECOND AUTHOR}
% 
% \address{Group, Laboratory, Address\\
% City, State ZIP/Zone, Country\\
% second\_author@group.com}

\maketitle

\begin{history}
\received{Day Month Year}
\revised{Day Month Year}
%\accepted{Day Month Year}
%\comby{(xxxxxxxxxx)}
\end{history}

\begin{abstract}

%The presence of asymptotic quantum features in the presence of dissipation is
%studied for different widely used indicators of quantumness, namely
%entanglement, quantum discord and sub-poissonian statistics. The system
%considered is a pair of coupled harmonic oscillators dissipating in separate or
%common thermal environments.

The dynamics of a pair of coupled harmonic oscillators in separate or common
thermal environments is studied, focusing on different indicators of
quantumness, such as entanglement, twin oscillators
correlations and quantum discord. We compare their decay under the effect of
dissipation and show, through a phase diagram, that entanglement is more likely to survive asymptotically
than twin oscillators correlations.

\end{abstract}

\keywords{Quantum correlations; Quantum statistical methods.}

\section{Introduction}
The characterization of correlations of a quantum state is object of an intense field of investigation, due to both its fundamental
scientific interest and its importance towards the implementation of quantum technologies.\cite{nielsen-chuang}
Entanglement has been traditionally considered as a fundamental resource to obtain quantum computational advantages, and has been used as
the main indicator of the quantumness of correlation. Indeed, as shown in Ref.~\refcite{linden}, for pure-state computation, exponential
speed-up only occurs if entanglement grows with the size of the system. Once mixed-state computation is considered, however, signatures of
quantum speed-up can come out using factorized states as, for instance, in the so-called Deterministic Quantum Computation with one Qubit
(DQC1).\cite{DQC1} Decoherence effects due to dissipation are known to be detrimental for entanglement that is indeed 
disappearing after a transient time (and not asymptotically).\cite{yu}

Among many other attempts to quantify quantum correlations, a predominant role has been assumed by quantum discord.\cite{zurek,henderson} It
has been introduced with the aim of capturing {\it all} quantum correlations, including entanglement. However, the relationship between
these two quantities is still unclear, since they seem to capture different properties of the states. 
In Ref.~\refcite{mdms} it is 
shown, for instance, that even if  discord and entanglement are the same for pure states, $mixed$ states maximizing the discord 
in a given range of classical correlations are actually separable. 
Recently, the analytical expression of
quantum discord has been obtained also for Gaussian states \cite{paris-adesso} opening the possibility to use it for continuous
variables.\cite{braunstein}

From a different point of view, the quantumness of a system
%, not necessarily based on the study of correlations, 
can be measured through other kinds of indicators, widely developed, for instance, in the field of quantum optics. 
%An example is the negativity of some normally
%ordered variance (for instance photon numbers) that can be associated to sub-poissonian statistics.
A well-known example are quantum correlations between two twin beams generated in optical
parametric oscillators.\cite{walls,loudon} The quantumness of the state of the emitted light is
measured by the absence of fluctuations in their intensities difference. This absence of noise 
is equivalent to  the negativity of this variance for normal ordered operators and was
first predicted by Reynaud and collaborators\cite{reynaud} and experimentally measured in Ref.~\refcite{heid}.
 Our aim in this work is to see if there is any connection between the latter
correlations, which we will call ``twin oscillators correlations'', with entanglement and discord, comparing their decaying and robustness.

In this paper, we will consider one of the most fundamental interacting systems, i.e.  two coupled
harmonic oscillators, in the presence of dissipation due to the interaction with a thermal
environment. Two extreme scenarios we are going to investigate are represented by the so-called
``common bath'', where the two oscillators are thought  to be so ``close'' with each other that they
interact with the same thermal modes, and the case of ``separate baths'', where the dephasing
channels are completely independent.\cite{klesse} It was recently shown\cite{liu,paz-roncaglia}
that two identical oscillators in the presence of a common bath can exhibit asymptotic entanglement
robust against decoherence, depending on the bath temperature and initial squeezing. This is a very
peculiar case and this behavior is generally lost if the two oscillators are not
identical.\cite{galvepra2010} Still, slow decay of {entanglement and robust} quantum correlations
appears in the presence of synchronization between detuned oscillators, as shown in
Ref.~\refcite{sync}. 

Studying the dynamics of the system through the master equation approach, we want both to analyze  the behavior of quantum correlations,
considering entanglement and quantum discord, and classify the global quantumness of the state using the variance of the difference of the
occupation numbers.  First, we will define the model and discuss its solution; afterward, the relevant indicators will be defined; as a
final step, we will study the  dynamics of such indicators.

\section{The model}

Let us consider two quantum harmonic oscillators allowing for diversity in their frequencies and direct coupling as well as dissipation
in thermal environment.\cite{petruccione} The model describing the system dipped into two identical but independent (separate)
thermal baths is given by the total $H=H_S+H_B+H_{SB}$, where the system Hamiltonian 
\begin{equation}
H_{S}=  \frac{p_1^2}{2}+\frac{1}{2}\omega_{1}^2x_1^2+ \frac{p_2^2}{2}+
\frac{1}{2}\omega_{2}^2x_2^2 +\lambda x_1 x_2
\end{equation}
describes two oscillators with frequencies $\omega_{1,2}$, unitary masses and
coupled through their positions,  
\begin{equation}
 H_{B}=\sum_{k}\sum_{i=1}^2\left(\frac{P_k^{(i)2}}{2}+\frac{1}{2}  
 \Omega_{k}^{(i)2}  X_{k}^{(i)2}\right)          
\end{equation}
 is the free Hamiltonian of two (identical) baths of harmonic oscillators (labeled by $k$), and 
\begin{equation}
H_{SB}=\sum_{k}\lambda_{k}^{(1)} X_{k}^{(1)}x_1+\sum_{k}\lambda_{k}^{(2)}
X_{k}^{(2)}x_2\label{sb}
\end{equation}
encompasses the system-bath interaction. 

The  case of a common bath is obtained by considering only $i=1$ in $H_{B}$  and
\begin{equation}
H_{CB}=\sum_{k}\lambda_{k} X_{k}(x_1+x_2)\label{cb}.
\end{equation}
The effective dissipation takes place therefore only in the sum of positions $(x_1+x_2)$, while in the case of separate baths both positions
$x_1$ and $x_2$ are
independently coupled with the thermal bath.

%by putting $X_{k}^{(1)}=X_{k}^{(2)}$, $P_{k}^{(1)}=P_{k}^{(2)}$,  and
%$\lambda_{k}^{(1)}=\lambda_{k}^{(2)}$.
The thermal bath is assumed here to be Ohmic, with a Lorentz-Drude
cut-off parameter $\Lambda$, and its spectral density\cite{petruccione} is
\begin{equation}\label{JOhm}
 J(\Omega)=\gamma_0 \omega \theta(\Lambda-\omega)
\end{equation} 
In the following we will consider the cut-off frequency always larger than any frequency involved in the oscillators free dynamics, that is
$\Lambda>>\omega_{1,2}$ and weak coupling $\gamma_0$. 

The analysis of the dissipation of identical oscillators in common and separate baths was given in  
Refs.~\refcite{liu,paz-roncaglia,paz-roncaglia2}. In these works it is shown that in presence of a common bath entanglement can persist
in the asymptotic state, in spite of dissipation. The master equations describing the time evolution of the reduced density matrix of
the system  (obtained by tracing out the degrees of freedom of the bath) when including the frequency diversity of the oscillators have been
reported in Refs.~\refcite{galvepra2010,sync} in the weak coupling limit (between system and environment) and without relying on the
rotating wave approximation.  Once the master equation has been obtained, we can explicitly write down and solve the equations of motion
for all the operators moments.  If the initial state of the two oscillators is Gaussian with vanishing average positions and
momenta, the complete information is contained in the matrix of second moments (the covariance matrix).
Detailed dynamical equations,  for the reduced density matrix and for the position and momenta second order moments, are given in 
 Ref.~\refcite{galvepra2010}, also in the system eigenmodes basis\cite{sync}
 and not reproduced here. The limit case of high temperatures in presence of a non-Markovian environment
 is studied in Ref.~\refcite{vasile}.

\section{Quantifiers of quantumness}

In this section, we briefly review the quantifiers we will use to characterize the time evolution of our system.

\subsection{Entanglement}

While, in general,  measures of entanglement have only been developed for pure states,
the case of Gaussian density matrices, together with the case of qubits, is one of the exceptions, since a necessary and sufficient criterion of separability exists.\cite{braunstein}

For  pure bipartite states $|\phi_{AB}\rangle=\sum_n c_n|u_n\rangle|v_n\rangle$, independently on their nature, entanglement can be calculated
through the von Neumann entropy (entropy of entanglement) of one of the two reduced density matrices: $E=-{\rm Tr}_A (|\phi_{AB}\rangle\langle\phi_{AB}|\log |\phi_{AB}\rangle\langle\phi_{AB}|)=-{\rm Tr}_B (|\phi_{AB}\rangle\langle\phi_{AB}|\log |\phi_{AB}\rangle\langle\phi_{AB}|)$. 
In the case of mixed states $\varrho$, however, the von Neumann entropy cannot be used since 
the mixedness of the reduced density matrices cannot discriminate between  entanglement and lack of purity of $\varrho$. 
A sufficient criterion (the so-called Peres-Horodecki criterion) for detecting entanglement can 
be obtained by considering the positivity of the partial transpose $\rho^{T_B}$ (or, equivalently, $\rho^{T_A}$), 
i.e. of the matrix obtained by only transposing the degrees of freedom of one of the two sub parties.\cite{peres,horodecki} 
Indeed, the presence of negative eigenvalues of $\rho^{T_B}$  witnesses that $\rho$ has not the form of a factorized density matrix.
As said before, in the case of Gaussian states, the separability of the partial transpose is also necessary to detect entanglement,\cite{simon}
and the modulus of the sum of the negative eigenvalues of $\rho^{T_B}$ (${\cal N}$) has been shown to be an entanglement monotone.\cite{vidal}

Since Gausssian states are completely characterized by their first and second moments, and first moments can be set to zero with local operations that do no modify entanglement,
the covariance matrix can  be used to check the positivity of the  partial transpose.
The logarithmic negativity, which represents an upper bound to the distillable entanglement, is defined as $E_{{\cal N}}=\log_{2}(2{\cal N}+1)$ and is related to the smallest symplectic
eigenvalue of the covariance matrix of $\rho^{T_B}$ ($\lambda_{-}$):  
\begin{equation}
 E_{{\cal N}}=\max[0,-\log_{2}2\lambda_{-}]. 
\end{equation}
In contrast to other entanglement measures, logarithmic negativity does not reduce to entropy of entanglement on pure states.

This entanglement quantifier was also considered in the context of coupled dissipative harmonic oscillators in the mentioned 
works.\cite{liu,paz-roncaglia,galvepra2010,sync,paz-roncaglia2,vasile}

\subsection{Quantum discord}

In classical information theory, the mutual information of a bipartite system 
can be calculated through two equivalent formulae related by Bayes rule:  
we have $\mathcal{I}(A:B)=\mathcal{J}(A:B)$, with  $\mathcal{I}(A:B)=H(A)+H(B)-H(A,B)$ 
and $\mathcal{J}(A:B)=H(A)-H(A|B)$, where $H(.)$ is the Shannon entropy and $H(A|B)$ 
is the conditional Shannon entropy of $A$ given $B$. 

The quantum counterparts of $\mathcal{I}(A:B)$ and $\mathcal{J}(A:B)$, however, 
differ substantially.\cite{zurek} By replacing the Shannon entropy with the von 
Neumann entropy of a given bipartite state $\varrho$ [$S(\varrho)=-{\rm Tr}\varrho\log_2\varrho$], we obtain the quantum mutual information
\begin{equation}
\mathcal{I}(\varrho)=S(\varrho_A)+S(\varrho_B)-S(\varrho),
\end{equation}
where $\varrho_{A(B)}$ are the reduced states after tracing out party $B(A)$. 
Due to the nature of measures in quantum mechanics, $\mathcal{J}(\varrho)$ depends on the measurement realized on $B$. 
Classical correlations are then  defined as \cite{henderson}
\begin{equation}
{\cal{J}}(\varrho)_{\{\Pi_j^B\}}=\min [S(\varrho_A)-S(A|\{\Pi_j^B\})],
\end{equation}
with the conditional entropy defined as $S(A|\{E_j^B\})=\sum_ip_iS(\varrho_{A|E_i^B})$, $p_i={\rm
Tr}_{AB}(E_i^B\varrho)$ and where $\varrho_{A|E_i^B}= E_i^B\varrho/{p_i} $ is the density
matrix after a positive operator valued measure (POVM) $(\{E_j^B\})$ has been performed on $B$.
Quantum discord is defined as the difference between ${\cal{I}}(\varrho)$ and ${\cal{J}}(\varrho)$:
\begin{equation}
\label{eqdisc}
\delta_{A:B}(\varrho)=\min_{\{E_i^B\}}\left[S(\varrho_B)-S(\varrho)+S(A|\{E_i^B\})\right].
\end{equation}
While the calculation of quantum discord, being based on a minimization procedure,
is in general an unsolved problem, in the case of Gaussian states, an analytical formula has been obtained.\cite{paris-adesso} 
This allowed to consider quantum discord also in the context of  continuous variable quantum information.\cite{braunstein}

The quantum discord measures in some sense how much disturbance is caused when
trying to know about party A when measuring party B, and has been shown to be
null only for a set of states with measure zero.\cite{acin}
It was shown to be a useful resource  in the DQC1 algorithm,\cite{DQC1} where the quantum
speed up does not rely on entanglement and, given its inequivalence to
entanglement (except for pure states), it hints at more general definitions
of what is quantum in a correlation.

The dynamics of quantum correlations, as quantified by the discord, and mutual information between quantum harmonic oscillators have been 
recently studied between different oscillators focusing on different parameters regimes and showing that the robustness (slow decay) of these
correlations is related to the presence of a synchronous dynamics.\cite{sync}

\subsection{Twin oscillators}

In the context of quantum optics, the discrimination between the predictions of classical
and quantum theories has been a wide field of investigation.
A tool to investigate the violation of classical inequalities is 
given by the variance of the difference of the occupation numbers:
\begin{equation}
d=\langle:\Delta^2(n_1-n_2):\rangle\label{vard}
\end{equation}
where, as usual, $n_i$ is the occupation number operator of each oscillator and $\langle:.:\rangle$
indicates normal ordering.  The quantumness of correlations in the occupation numbers derives from the
absence of noise when subtracting the oscillators intensity fluctuations. This is equivalent (in normal ordering) to the
negativity of the variance $d$ and is a consequence of the negativity of the Glauber-Sudarshan
quasi-probability $P$. 

We note that for identical oscillators Eq. \ref{vard} is identical to 
$\langle: (n_1-n_2)^2:\rangle$ and this indicator characterize anti-bunching, being a cross correlation
larger than the autocorrelation. When the system or the state are not symmetrical in the two components,
the negativity of the latter second order moment would be one of many possible quantum indicators,
implying negativity of the correspondent $P$ distribution not associated to anti-bunching.\cite{lee} The existence of these strong
correlations in optics generally comes from the simultaneous generation of pairs of photons in nonlinear
processes, and this generally characterizes twin beams.\cite{loudon} This is a rather robust phenomenon in
complex spatiotemporal dynamics.\cite{epjd}

In the following we will study the temporal dynamics of $d$, that we will name twin oscillator
correlations. Given the Gaussian character of the 
initial state we want to study, the fourth order moments can be obtained from the covariance
matrix.\cite{gardiner} The dynamics of this indicator was already considered in Ref.~\refcite{galvepra2010}
in comparison with entanglement through few examples, showing a similar decay after a finite time
transient. However, the possibility to get asymptotically twin oscillators was not considered there and
will be addressed in the next section where we will also fully analyze the role of the  squeezing of the
initial state and of the temperature. On the other hand, this indicator ($d$) has been considered in 
Ref.~\refcite{vasile} in a different regime, for high temperatures focusing on the short time decay   in
presence of non-Markovian environment.

\section{Correlations dynamics}

Let us consider an initial two-mode squeezed state 
 \begin{eqnarray}\label{TMS}
| \Psi_{TMS}
\rangle=\sqrt{1-\mu}\sum_{n=0}^\infty \mu^{n/2}|n\rangle|n\rangle, 
 \end{eqnarray}
where $\mu=\tanh^2 r$ and $r$ is the squeezing amplitude. We know that for this state $d=-2\mu/(1-\mu)<0$.
 We want to study how the different indicators dynamically behave considering three different scenarios:
(i) the case of different frequencies for a common bath;
(ii) the case of equal frequencies for separate baths;
(iii) the case of equal frequencies ($\omega_{1}=\omega_{2}$) for a common bath.    %Furthermore, we are interested to see how the temperature affects the robustness 
As shown in Refs.~\refcite{liu} and ~\refcite{paz-roncaglia}, for equal frequencies, in the case of a common environment, the asymptotic state is expected
to be entangled. This is due to the  fact that one of the degrees of freedom of the system [the mode $x_-=(x_1-x_2)/\sqrt{2}$] is actually
frozen, since it does not interact with the bath, and represents a decoherence-free subspace. Moving away from the resonance condition, a
full thermalization process takes place, and the final (Gibbs)  state can be entangled only in the very low temperature regime.  Motivated
by these results, we then want to investigate whether the existence  of this noiseless channel also protects other aspects of the
quantumness of the state under evolution.  We start $\lambda=0$ (no direct coupling between the oscillators) to avoid fast oscillations and 
$\omega_1$ as a scale unit of measure of energy.%, scaling all other quantities with it. 
Then, we choose for the system bath coupling  $\gamma=2\times 10^{-2}/\pi \omega_1$ 
(weak coupling limit) 
with cut-off $\Lambda=20\omega_1$, and squeezing $r=2$ in the initial state (\ref{TMS}). 

\begin{figure}[pb]\label{figura2}
\centerline{\psfig{file=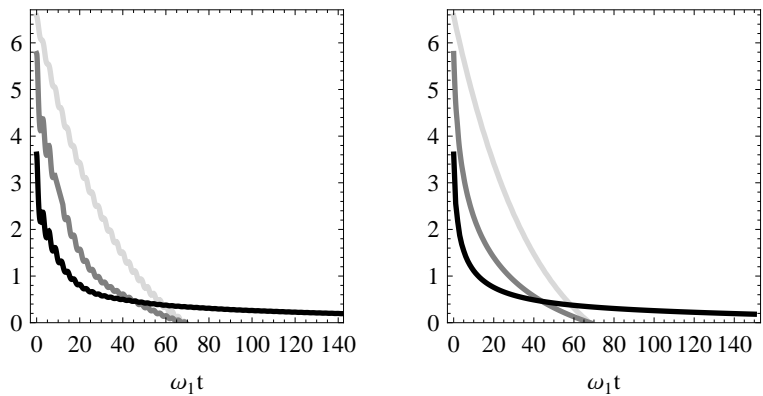}} 
\vspace*{8pt}
\caption{Dynamics of logarithmic negativity (gray), quantum discord (black), and $\max[0,-d/4]$ (light gray) for $T=\omega_{1}$.
Left panel: $\omega_{2}=1.2\omega_{1}$ and common bath; right panel: $\omega_{2}=\omega_{1}$ and separate baths.}
\end{figure}

In Fig.~1, we show the dynamics  of logarithmic negativity, quantum
discord and $\max[0,-d/4]$ ($d$ is scaled by a factor 4 for the  sake of comparison). We
start considering resonant oscillators ($\omega_{1}=\omega_{2}$) coupled to  separate baths
and detuned oscillators  ($\omega_{1}\neq\omega_{2}$) coupled to a common environment.
Entanglement and $\max[0,-d/4]$ vanish in a finite time,\cite{galvepra2010}
being more fragile than quantum
discord, which is exponentially decaying.\cite{sync} As a matter of fact, it is known that states with
vanishing discord are rare,\cite{acin,mdms} and in the presence of dissipation this indicator
does not  experience a sudden death process. 

The case of equal frequencies and a common environment is that one showing entanglement in
the asymptotic state.\cite{liu,paz-roncaglia}  As we can observe from the right panel of
Fig.~2, for temperature $T=\omega_{1}$, both entanglement and quantum
discord reach a stationary regime after a transient phase, while the negativity of $d$
disappears in a finite time. By lowering $T$ up to $0.1\omega_{1}$, i.e. in a
regime where quantum effects are stronger, we see that the asymptotic value of entanglement
and discord is increased, but we also observe that the variance of the difference between the
occupation numbers becomes negative for infinite time. 
% In Fig.~3  we plot the three indicators for larger temperature $T=10\omega_{1}$, 
% $\omega_{2}=1.2\omega_{2}$, $\lambda=0.2\omega_{1}^2$, $r=0.5$ and common
% bath. All quantum markers between the states are degraded even in presence of a non vanishing
% coupling.
In Fig.~3,  we reproduce the left panel of Fig.~2 by adding a finite direct coupling between the two oscillators $\lambda=0.2\omega_{1}^2$.
While the presence of $\lambda$ induces fast oscillations in  the three observables, all of them are still present in the asymptotic regime.

Finally, comparing these first figures, it seems then that the sub-poissonian
character of fluctuations ($d<0$) is rather fragile and that oscillators remain twin
asymptotically only at very low temperature.

\begin{figure}[pb]\label{figura1}
\centerline{\psfig{file=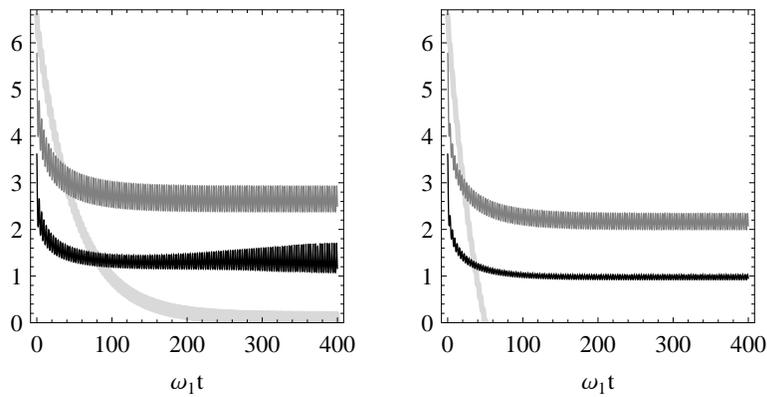}} 
\vspace*{8pt}
\caption{Dynamics of logarithmic negativity (gray), quantum discord (black), and $\max[0,-d/4]$ (light gray) (see the main text for details)
for $\omega_{1}=\omega_{2}$ and a common bath. Left panel: $T=0.1\omega_{1}$; right panel: $T=\omega_{1}$.}
\end{figure}

\begin{figure}[width=5cm]\label{figura3}
\centerline{\psfig{file=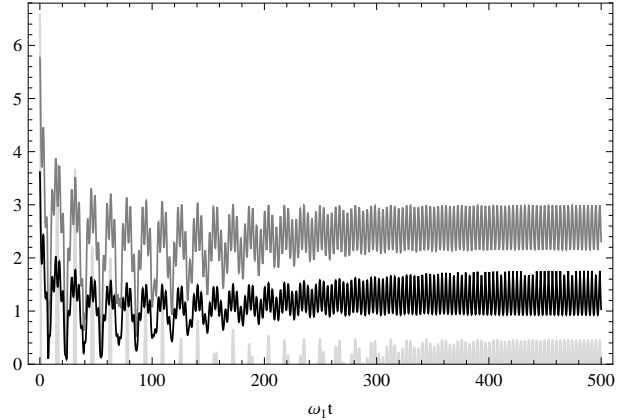,width=8cm}}
\vspace*{8pt}
\caption{Dynamics of logarithmic negativity (gray), quantum discord (black), and $\max[0,-d/4]$ (light gray)  
for a common bath.
Here, $\omega_{2}=\omega_{1}$, $T=0.1\omega_{1}$ and $\lambda=0.2\omega_{1}^2$.}
\end{figure}

\section{Asymptotic entanglement and twin oscillators correlations}

In order to have a general view of the persistence of both entanglement and twin oscillators
correlations for a common bath, we consider the respective asymptotic states. The asymptotic entanglement for
identical decoupled oscillators dissipating in a common bath was given in   Ref.~\refcite{paz-roncaglia}
as a function of the temperature and the squeezing of the initial state.  We show the phase diagram for
presence and absence of asymptotic entanglement in Fig.~4, in the weak coupling limit. 

We then estimate $d$ in the asymptotic limit
($t\rightarrow \infty$), assuming thermalization for the mode $x_+=(x_1+x_2)/\sqrt{2}$ and free dynamics
for the mode $x_-=(x_1-x_2)/\sqrt{2}$ depending on the
initial state. By explicitly writing down $d$ and by replacing all the entries of the
two-oscillator  covariance matrix by their asymptotic value, we  conclude that the variance of the
difference between the occupation numbers is negative when the following inequality is satisfied:
\begin{equation}
 \langle x_-^2\rangle( 2\langle x_+^2\rangle-1)+\langle p_-^2\rangle(2 \langle p_+^2\rangle-1)+1<\langle x_+^2\rangle+\langle p_+^2\rangle.
\end{equation}
Since  the observables of the  non dissipating mode $x_-$ are oscillating in time, in order to find
negative values for $d$, we will take the minimum in a period.  It follows then, as for entanglement,
it is equivalent to start from a two-mode squeezed state or two (separable) squeezed states. Finally we
obtain the phase diagram represented in Fig.~4, where entanglement and negative $d$
(or negative $P$ distribution) are obtained below the corresponding lines.
We see that twin oscillators correlations are achieved only for very low temperatures, 
in contrast with  entanglement  for
which the detrimental effect of temperature can be compensated by stronger initial squeezing. 
As a matter of fact, for $r\gtrsim 1$ there are no twin oscillators unless the temperature is very low ($T\lesssim 0.25 \omega_1$).
This critical temperature corresponds to $2\langle x_+^2\rangle=1$.

An important remark is that the threshold curve for $d<0$ is not continuous. In fact, for $r=0$ we have $d=0$ for any $T$. On the other hand, 
for $r\to 0^+$, the presence of twin correlations is determined by the sign of  $\langle x_+^2\rangle-\langle p_+^2\rangle$, and they only appear 
below a finite critical temperature.

\begin{figure}[width=3cm]\label{figura4}
\centerline{\psfig{file=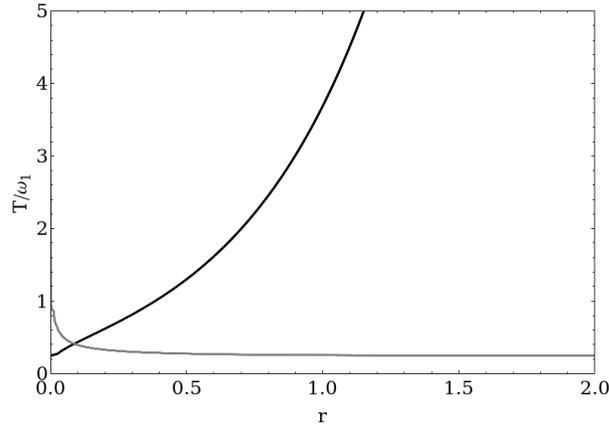,width=8cm}}
%\centerline{\psfig{file=fig4.eps}}
\vspace*{8pt}
\caption{Phase diagram for entanglement (logarithmic negativity) and twin correlations ($d<0$)
as a function of the temperature ($T/\omega_{1}$) and initial squeezing, in the weak coupling limit 
($\gamma=2\times 10^{-2}/\pi \omega_1$ )
and for identical  ($\omega_{1}=\omega_{2}$) and decoupled ($\lambda=0$) oscillators.
The asymptotic state is entangled or displays twin oscillators correlations in the lower areas, as limited by black and gray lines respectively.}
\end{figure}

From these results, we learn that the presence of a decoherence-free channel allows the preservation of
the quantum character of the state,  which manifests itself both through the presence of correlations
(entanglement and discord) and the negativity of $d$.  This latter characteristic, signature of a
negativity of the $P$ distribution, is more
fragile than the other quantum markers considered here.

Entanglement and twin correlations have been analyzed also in Ref.~\refcite{paris} in a different
case, for a pair of optical field modes obtained from parametric down-conversion when the input
light is in a thermal and separable state. Even if the states obtained from (different processes of)
thermalization of a squeezed state, as  considered here, and the thermal state after unitary action of parametric
down conversion, as in Ref.~\refcite{paris},  are different, we see that in both cases entanglement
is more likely to be found that twin oscillators correlations.

\section{Conclusions}

We have studied the effects of dissipation on the quantum character of the state of two coupled
harmonic oscillators. We discussed the system quantumness  considering the dynamics of different
indicators, such as  entanglement, quantum discord, and twin oscillators correlations, a signature of
negative values of the $P$ distribution.  We find that in general all these quantumness indicators
vanish  in the asymptotic limit of large times when they reach the equilibrium (apart from an
exponentially small value of  discord) unless a decoherence-free channel exists. In our model this
channel is represented by the coupling of the two resonant (identical) oscillators to a common
environment. Still, out of resonance between the two oscillators, it would possible, in line of
principle, to restore the noiseless channel by unbalancing the coupling of the oscillators to the
bath.\cite{galvepra2010}  We have also found that, whenever entanglement cannot be asymptotically
preserved, its death time  becomes similar to that of   the variance of the difference between the
occupation numbers ($d$).

After showing the effect of detuning between oscillators, (common or separate) environment,
oscillators coupling and temperature through few examples, we obtained the phase diagram for asymptotic
entanglement and twin oscillators correlations. In the case of identical decoupled oscillators, we have
found that in order to find a negative value of $d$ at very long time,  the temperature needs to be
smaller than the one allowing for asymptotic conservation of entanglement.
Moreover, the effect of temperature on entanglement can be compensated by stronger initial squeezing, 
being this not the case for twin oscillators correlations. 
 So, this quantifier ($d<0$) is
even more fragile under dissipation than the entanglement itself. This observation could be more general as
it is similar to what  found in a rather different system in Ref.~\refcite{paris}.

This asymptotic phase diagram has been obtained for identical oscillators and a common thermal bath. 
Once the frozen degree of freedom disappears (either by detuning the oscillators or by introducing
separate baths), the system is expected to asymptotically thermalize in all degree of freedom. Its
quantum character is then generally lost, unless other mechanisms are introduced, as for instance
driving the system out of equilibrium.\cite{fernando} As a matter of fact, twin oscillators
correlations are mostly related to the presence and the direction of squeezing of the damped eigenmode 
(that is the sign of $\langle x_+^2\rangle-\langle p_+^2\rangle$).

\section*{Acknowledgments}
Funding from FISICOS (FIS2007-60327), CoQuSys (200450E566), 
"Accion Especial" CAIB (AAEE0113/09) projects are acknowledged. GLG is supported by the Spanish Ministry of Science and Innovation 
through the program ``Juan de la Cierva''. 
FG is supported by the CSIC through the program ``Junta para la Ampliaci\'on de Estudios''

\end{document}